\documentclass[final,5p,number,sort&compress,times]{elsarticle}

\usepackage{amssymb}
\usepackage{graphicx}
\usepackage{color}
\usepackage[vertfit]{breakurl}
\usepackage[caption=false, labelformat=empty]{subfig}
\usepackage{microtype}
\usepackage{comment}

\journal{Nuclear Instruments and Methods in Physics Research Section A}

\begin{document}

\begin{frontmatter}



\title{Low energy neutron background in deep underground laboratories}

\author[lngs,und]{Andreas Best \corref{corre}}
\author[und]{Joachim G\"{o}rres}
\author[lngs]{Matthias Junker}
\author[mpi]{Karl-Ludwig Kratz}
\author[lngs]{Matthias Laubenstein}
\author[und]{Alexander Long}
\author[lngs]{Stefano Nisi}
\author[und]{Karl Smith\fnref{utx}}
\author[und]{Michael Wiescher}

\address[lngs]{INFN, Laboratori Nazionali del Gran Sasso (LNGS), 67010 Assergi, Italy}
\address[und]{Department of Physics and The Joint Institute for Nuclear Astrophysics, University of Notre Dame, Notre Dame, IN 46556, United States}
\address[mpi]{Department for Biogeochemistry, Max-Planck-Institute for Chemistry, 55020 Mainz, Germany}

\cortext[corre]{Corresponding author. Fax: +39 0862 437 570. \emph{E-mail address:} andreas.best@lngs.infn.it (A. Best)}
\fntext[utx]{Present address: The University of Tennessee, Knoxville, Knoxville, TN 37996, United States}

\begin{abstract}
The natural neutron background influences the maximum achievable sensitivity in most deep underground nuclear, astroparticle and double-beta decay physics experiments. Reliable neutron flux numbers are an important ingredient in the design of the shielding of new large-scale experiments as well as in the analysis of experimental data.

Using a portable setup of $^3$He counters we measured the thermal neutron flux at the Kimballton Underground Research Facility, the Soudan Underground Laboratory, on the 4100 ft and the 4850 ft levels of the Sanford Underground Research Facility, at the Waste Isolation Pilot Plant and at the Gran Sasso National Laboratory. Absolute neutron fluxes at these laboratories are presented.
\end{abstract}

\begin{keyword}
Neutron flux \sep Underground laboratory \sep He-3 counters


\end{keyword}

\end{frontmatter}


\section{Introduction}
\label{sec:intro}
The major challenge of many modern physics experiments is the measurement of very small event rates (down to a few events per year, e.g. \cite{Auger:2012}). Locating the equipment
deep underground is a first step towards achieving this goal, since the rock overburden shields the experiment from cosmic rays. Generally the  muon flux and the secondary gamma- and neutron
fluxes are attenuated by a few orders of magnitude compared to the surface \cite{Bettini:2012}.

A wide range of deep underground experiments are sensitive to background neutrons: In underground nuclear astrophysics stellar neutron sources need to be
measured down to very low cross sections. Elastic scattering of neutrons can mimic signals expected from WIMP interactions. Neutrinoless $\beta\beta$ decay searches can
be influenced by $\gamma$ rays emitted after neutron inelastic scattering or capture and also by the decay of unstable nuclei produced through neutron capture. Therefore,
the neutron background needs to be understood and sufficient shielding needs to be implemented to limit its impact on the experimental sensitivity.

The underground neutron flux is mostly due to spontaneous fission of $^{238}$U in the cavity walls, $(\alpha,n)$ reactions induced by $\alpha$-particles from the natural
radioactivity of the underground environment and from the activity of the experimental setup itself \cite{HashemiNezhad:1995, Balbes:1997, HashemiNezhad:1998, Chazal:1998, Carmona:2004, Kim:2004, Wulandari:2004, Tziaferi:2007, Mei:2010}. The cosmic-ray induced
neutron flux is two to three orders of magnitude lower than the radiogenic component \cite{Mei:2006}.  Usually the laboratory neutron flux is simulated based
on the composition of the rock and the concentration of radioactive isotopes \cite{Tomasello:2008, Mei:2010}. However, these simulations carry a large degree of uncertainty and should be tested against measurements, if available.

Geological composition of the rock and the varying contents of uranium and thorium in the underground environment as well as differences in the water content of the surrounding rocks cause variations in the background
 between underground laboratories. In addition it has been found that even local differences in the composition of the rock can lead to background levels that vary by an
 order of magnitude between locations in the same laboratory (e.g., Halls A and C in Gran Sasso) \cite{Wulandari:2004}.
 
Some data on measured neutron backgrounds are available, but a direct comparison is made difficult by the variety of detection setups used and differences
 in the covered neutron energy range. In this work we present measurements of the thermal neutron fluxes at various underground locations using a portable setup
 of $^3$He detectors. Measurements
were done at the Waste Isolation Pilot Plant (WIPP) near Carlsbad, New Mexico; the Soudan Underground Laboratory in Minnesota; the Kimballton Underground Research Facility (KURF) in West Virginia; the Sanford
Underground Research Facility (SURF) located in the Black Hills in South Dakota and the Laboratori Nazionali del Gran Sasso (LNGS) in Italy.

\section{Site background parameters}
\begin{table*}[tb]
	\centering
	\caption{Properties of the visited underground sites. SURF values are stated for the 4850 foot level. See text for details on the WIPP U and Th concentrations.}
	\begin{tabular}{cccccc}
		\hline \hline 
        & WIPP \cite{Esch:2005}	& Soudan \cite{Ruddick:1996,Mei:2006}	& KURF \cite{Xu:2012}	& SURF \cite{Mei:2006,Mei:2010} & LNGS \cite{Ahlen:1990,Wulandari:2004}\\
\hline
Environment								& Salt				& ``Ely Greenstone'' & Limestone	& Poorman foundation & Limestone\\
Depth [m]								& 655				& 780		& 500		& 1500 & 1400\\
Equivalent depth [mwe]					& 2000				& 2090		& 1400		& 4300 & 3800\\
muon flux [$10^{-7}$ s$^{-1}$ cm$^{-2}$]& $4.77 \pm 0.09$	& $2.0 \pm 0.2$	& $\approx 20$	&	$0.044 \pm 0.001$ & $0.32 \pm 0.01$\\
$^{238}$U [ppm]                         & (0.48 -- 1.49) $\cdot 10^{-3}$ & 0.17 & & 3.4 & 6.8 (Hall A)\\
$^{232}$Th [ppm]                        & (1.01 -- 1.9) $\cdot 10^{-3}$ & 0.89 & & 7.1 & 2.2 (Hall A)\\
\hline \hline
	\end{tabular}
	\label{tab:site-overview}
\end{table*}

The measurements presented here cover very different background environments: WIPP is located in a salt formation, Soudan is a former iron mine, KURF is situated in an active limestone
mine, SURF is located in a retired gold mine; the Gran Sasso laboratory is in a limestone formation. The properties of each site relevant to the background conditions and measured
values of the muon flux are listed in table \ref{tab:site-overview}. Although WIPP is at a relatively shallow depth, the radiogenic radioactivity is very low due to the low-activity salt
environment. U and Th concentrations displayed in the table were measured by inductively coupled plasma mass spectrometry on salt samples. The low and high values from the range given
in table \ref{tab:site-overview} were determined on a clear and a rocky sample showing a slight coloration, respectively. Additional measurements by gamma-spectroscopy are in agreement with
these values and, by measuring the gamma-active daughter nuclides of the uranium and thorium decay chains, confirm that
they are in secular equilibrium.

The thermal neutron flux at WIPP has been reported previously as $(1.3 \pm 0.3) \cdot 10^{-7}$ cm$^{-2}$ s$^{-1}$ \cite{Balbes:1997}. At LNGS various measurements have been done;
none of them agree with each other, possibly due to a variation in the background depending on the location in the laboratory or due to unknown systematic uncertainties.
The reported values are: $(2.05 \pm 0.06
) \cdot 10^{-6}$ cm$^{-2}$ s$^{-1}$ \cite{Rindi:1988}, $(1.08 \pm 0.02) \cdot 10^{-6}$ cm$^{-2}$ s$^{-1}$ \cite{Belli:1989}, and $(5.4 \pm 1.3) \cdot 10^{-7}$ cm$^{-2}$ s$^{-1}$ \cite{Debicki:2009}.
Only higher-energy neutron data are available at the other sites.

\section{Experimental setup}
\subsection{Neutron detection with $^{3}He$ proportional counters}\label{sec:counters}
\begin{figure}[htb]
	\centering
	\includegraphics[width=\linewidth]{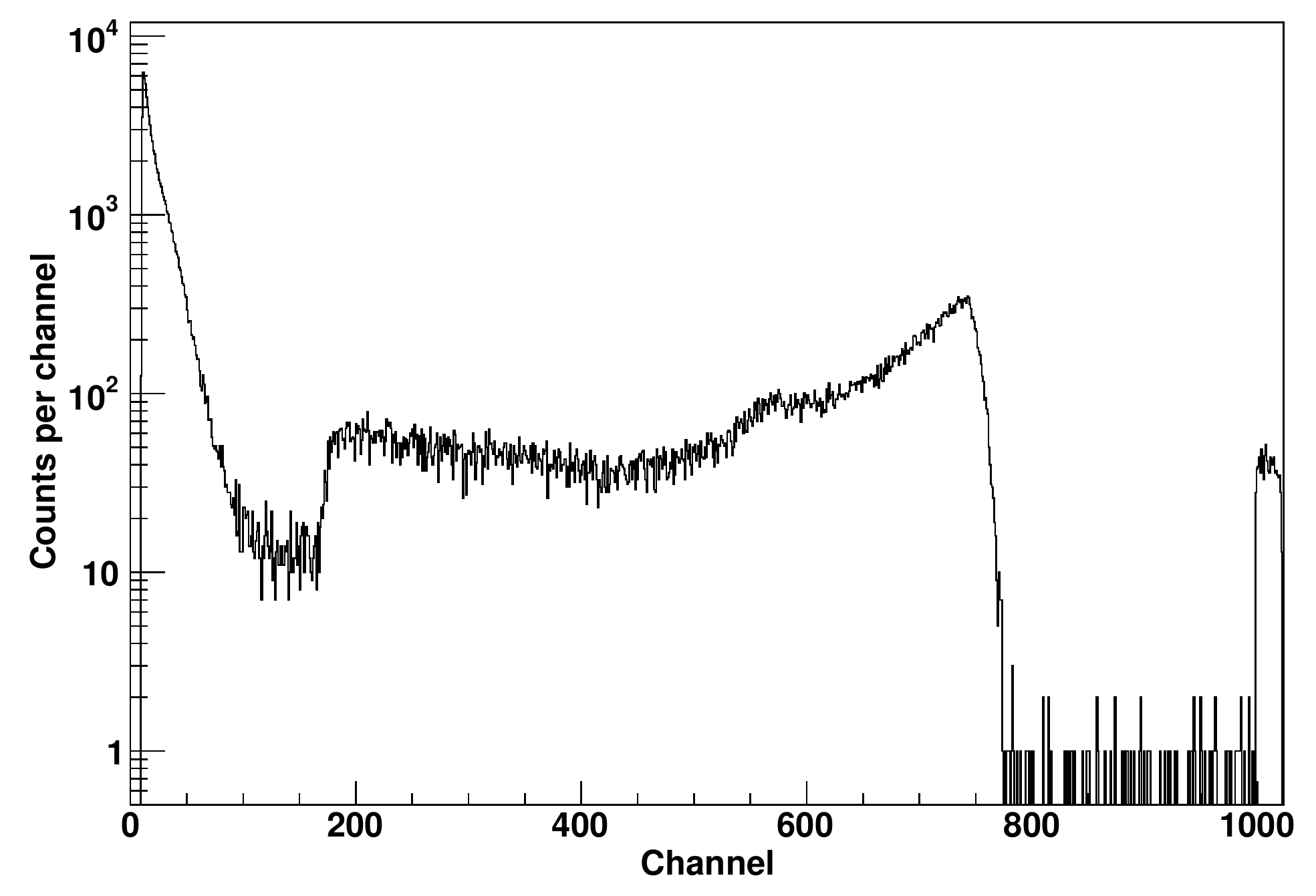}
	\caption{Typical spectrum from a $^3$He counter (1 channel $\approx$ 1 keV). A neutron generates a signal between channels $\sim$200 and 800. Signals above and below this region are due to alpha particles and electronic noise, respectively.}
	\label{fig:example-spec}
\end{figure}
The $^3$He detectors used in our experiments consist of an Al tube that is filled with $^3$He (and a small amount of CO$_2$ as quench gas) under a pressure of 10 bar. A wire
at a +1400 V potential runs through the center of the Al cylinder. $^3$He has a very high cross section for capturing thermal neutrons through the reaction $^3$He$(n,p)^3$H 
($\sigma$ = 5330 barn, Q = 764 keV \cite{Mughabghab:1981}). After a neutron has been captured the proton ($p$) and the triton ($T$) deposit their kinetic energy
(E$_p$ = 573 keV, E$_T$ = 191 keV) in the $^3$He gas. In case both reaction products are fully stopped in the sensitive volume a pulse with a height proportional to
764 keV is generated. Due to the finite volume of the detectors there is a chance that one or both nuclei hit the Al cylinder and only deposit a fraction of their energy in
the detector (wall effect). This gives lead to the characteristic pulse height spectrum of a $^3$He proportional counter. Fig. \ref{fig:example-spec} shows a typical spectrum 
taken with one of the $^{3}$He counters used in this work, with the two wall-effect peaks around channels 200 and 550. $\gamma$-rays and electronic noise generate counts 
below channel 180 and can be clearly distinguished from the neutron signals. 

\subsection{Intrinsic background}
\begin{figure}[h]
	\centering
	\includegraphics[width=\columnwidth]{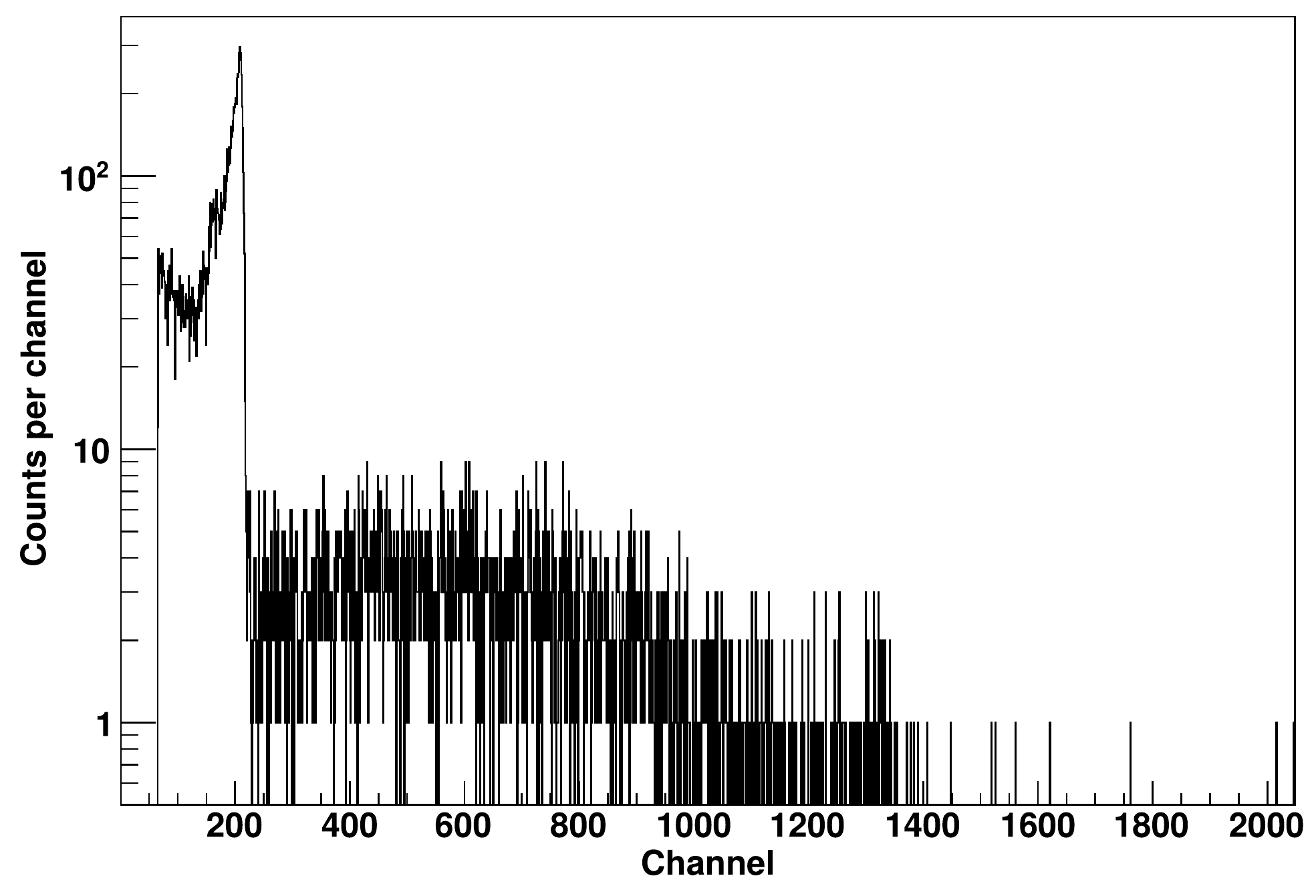}
	\caption{Low-gain spectrum. The thermal neutron peak is visible around channel 200. The counts at higher energies are due to the internal radioactivity of the counter itself.}
	\label{fig:alphas}
\end{figure}
The counts visible at energies above the neutron peak are due to the internal radioactivity of the counters themselves. $\alpha$-particles from the decay of the uranium and thorium
present in the detector walls generate background signals covering the area of the thermal neutron peak (at 764 keV and below) and extending up to 9 MeV \cite{HashemiNezhad:1998}. An additional background
component is due to microdischarges near the central wire of the detectors \cite{Langford:2013}. The count rate in this range orresponds to the sensitivity limit for low-level neutron detection.
The combined effect can be seen in Fig. \ref{fig:alphas}, where the amplification of the detector signal has been lowered to cover a wider energy range. The thermal neutron full-energy peak
lies at channel 200 with the internal detector background extending beyond it. This background is usually not of concern on the earth's surface but at the very low neutron background conditions in an
underground environment it becomes a major background component that can be stronger than the actual neutron signals. The average background rate integrated over the region of the neutron
peak is about $10^{-3} $s$^{-1}$ for the counters used here. Using the ratio of alphas in the neutron signal range to total alpha particles from the simulation and the area of the counter
one obtains a total alpha activity of the counters of about $6 \cdot 10^{-5} $cm$^{-2} $s$^{-1}$. This value is in very good agreement with measurements of similar He-3 counters and an assay of
commercially available aliuminum \cite{Langford:2013}.

\begin{figure}[htb]
	\centering
	\includegraphics[width=\columnwidth]{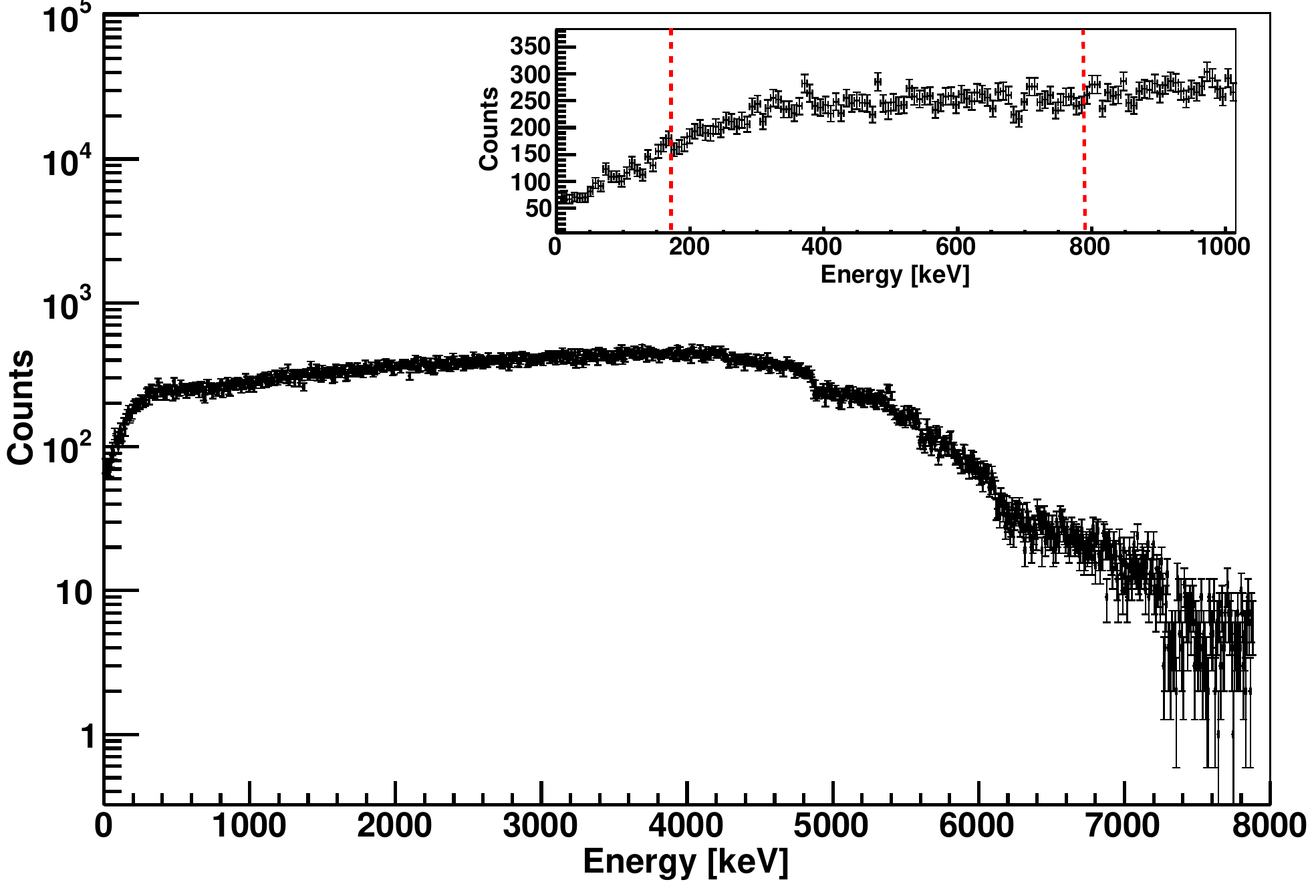}
	\caption{{\sc Geant4} simulation of the pulse height spectrum due to $\alpha$-particles emitted from the walls of a $^3$He counter. The region of the thermal neutron peak is marked by the dashed lines in the inset.}
	\label{fig:he3-alpha-sim}
\end{figure}

To model the $\alpha$-induced background {\sc Geant4} simulations \cite{Agostinelli:2003} were performed with $\alpha$-particles being emitted from inside the aluminum
container of the $^3$He counter. Contributions from decays in the anode material are negligible due to its much smaller surface area. Secular equilibrium in both the thorium and uranium
chains was assumed and the particle energies were randomly chosen from the $\alpha$-energies in the chains. Since the U and Th content of the containers is unknown various
simulations with different Th/U ratios were performed; the resulting spectra are not significantly different from each other. A spectrum for Th/U = 1 is shown
in Fig. \ref{fig:he3-alpha-sim}, the inset is a zoom into the energy region of the thermal neutron peak between 200 keV and 800 keV. In the area of interest the
energy deposition from alphas is featureless and the $\alpha$-induced background can be reliably subtracted from the underground neutron data.

\subsection{Detector components and response function}
The experimental setup used for the underground emasurements consisted of two $^{3}$He counters\footnote{The counters are 25.4 cm long (defined by the active part of the $^3$He inside the aluminum enclosure) and have a radius of 1.27 cm.} 
as described in Sec. \ref{sec:counters}.

{\sc Geant4}\footnote{Version 9.6.2, with the high precision neutron data libraries G4NDL4.2, including thermal cross sections.}
simulations resulted in a thermal neutron detection efficiency of 80 \%.

Details of the measurements at each site are described in the following section. In all cases the data were saved every hour so that the count rate could be checked for inconsistencies.

\section{Overview of the measurements the different sites}
\subsection{KURF}
Data at KURF were taken for approximately one month. The detectors were deployed in the experimental hall on the 14th level of the mine. At that depth the rock
overburden is approximately 500 meters, equivalent to 1400 meters of water. Since the laboratory is located in an active limestone mine the study had to take into account nightly blast from the
mining operations. No effect on the recorded count rate was observed.

\subsection{Soudan}
The experimental halls of the Soudan laboratory are located at a water equivalent depth of 2090 m (780 m below the surface) on the 27th level of the former Soudan Iron Mine.
The detectors were set up on the second floor of the Soudan 2 / CDMS II cavern. Data were taken over a period of $\sim$ 15 weeks. The detectors were calibrated with a $^{252}$Cf source.

\subsection{SURF}
Measurements at SURF were done on the 4100 foot and at two locations on the 4850 foot level. Three months worth of data were accumulated at 4100 feet. There the detector was
placed inside an airlock along the main drift between the Yates and Ross shafts. The first measurement at 4850 feet was done for two months in an alcove just outside the entrance to
the LUX and Majorana clean area. Then the detector was moved to the Majorana electroforming laboratory (``Temporary Clean Room'', TCR) and another 1.5 months of data were taken.
After the underground measurements were completed the setup was brought to the surface and a short calibration run using the natural neutron background was performed.

\subsection{WIPP}
The neutron flux at WIPP was measured for approximately five months. The detectors were located in a connex in the ``Q'' alcove near the air intake shaft.
WIPP lies at a depth of $\sim$ 655 m ($\approx 2000$ m.w.e.), roughly in the center of a 600 m thick salt formation. A distance of at least 600 m separates the low-level transuranic
waste storage areas from the experimental alcove, so that no adverse effect on the measured neutron flux is to be expected.

\subsection{LNGS}
The setup was located in the upper connex of the LUNA II experiment, situated in the ``interferometer tunnel'' just outside of Hall A. The averaged rock overburden
of LNGS is 1400 m (3800 m.w.e.) below the Gran Sasso mountain massif. The dominant rock is limestone. Before the measurement the detector calibration was checked using an $^{241}$Am source.

\begin{figure*}[htb!]
    \centering
    \includegraphics[width=\textwidth]{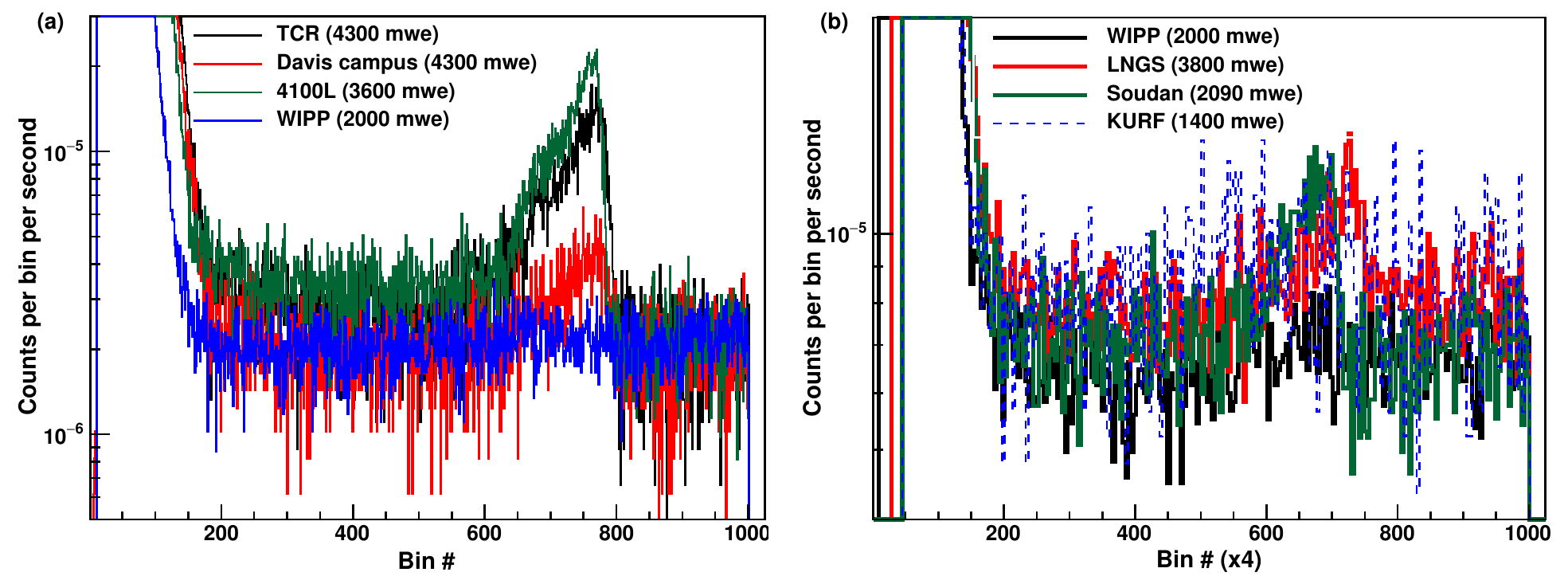}
    \caption{(color online) Spectra of the different measurements. (a) shows the results of the different SURF measurements in comparison with WIPP. (b) compares the LNGS, Soudan and KURF data to WIPP. The histograms in (b) were rebinned to easier differentiate them from each other.}
    \label{fig:spectra}
\end{figure*}

\section{Analysis and results}
\begin{table*}[htb!]
	\centering
	\caption{Experimental results. The upper limits are 2 $\sigma$ values, uncertainties are 1 $\sigma$.}
	\begin{tabular}{ccccc}
		\hline \hline 
		& WIPP	& Soudan & KURF	& LNGS \\
\hline
Measurement time [$10^6$ s]		&   12.7    & 8.9           & 2.3    & 8.4   \\
Neutron flux $[10^{-6} $cm$^{-2}$ s$^{-1}] (\pm stat. \pm sys.)$ &   $< 0.06$& $0.7 \pm 0.08 \pm 0.05$ & $ < 0.4$ & $0.32 \pm 0.09 \pm 0.02$ \\
\hline
		& SURF (4100L)	& SURF (Davis) & SURF (TCR) &\\
\hline
Measurement time [$10^6$ s]		&   6.4  & 4.9  & 4.5 & \\
Neutron flux $[10^{-6} $cm$^{-2}$ s$^{-1}] (\pm stat. \pm sys.)$ & $9.9 \pm 0.1 \pm 0.7$ & $1.7 \pm 0.1 \pm 0.1$ & $8.1 \pm 0.1 \pm 0.6$ & \\
\hline \hline
	\end{tabular}
	\label{tab:results}
\end{table*}
The thermal neutron flux $\phi$ can be calculated from the raw count rate $C_{tot.}$ in the energy region of the neutron peak, the expected background count rate $C_{bg}$, the detector
efficiency $\eta$ and its surface area $A = 213$ cm$^2$:
\begin{equation}
    \phi = \frac{C_{tot.}-C_{bg}}{\eta A}
    \label{eq:flux}
\end{equation}
Figure \ref{fig:spectra} shows the raw neutron data. To calculate the signal count due to neutrons the alpha counts
from above the neutron peak were extrapolated to lower energies and subtracted from the raw counts since those resemble a mixture of neutron, alpha (and microdischarge) signals. The flux at WIPP is so low that
even after 5 months of data taking no neutron signal is present. The spectrum therefore resembles pure background and can be used as a baseline level for comparison.

The thermal neutron fluxes at each site are listed in Table \ref{fig:spectra}. For WIPP and KURF it is only possible to determine upper limits (2 $\sigma$ limits are given). 
Systematic uncertainties of $\pm 5 \%$ each for the efficiency simulation and the extrapolation of the alpha background into the signal region are included in the table.

\section{Discussion}
\begin{figure}[htb!]
    \centering
    \includegraphics[width=\columnwidth]{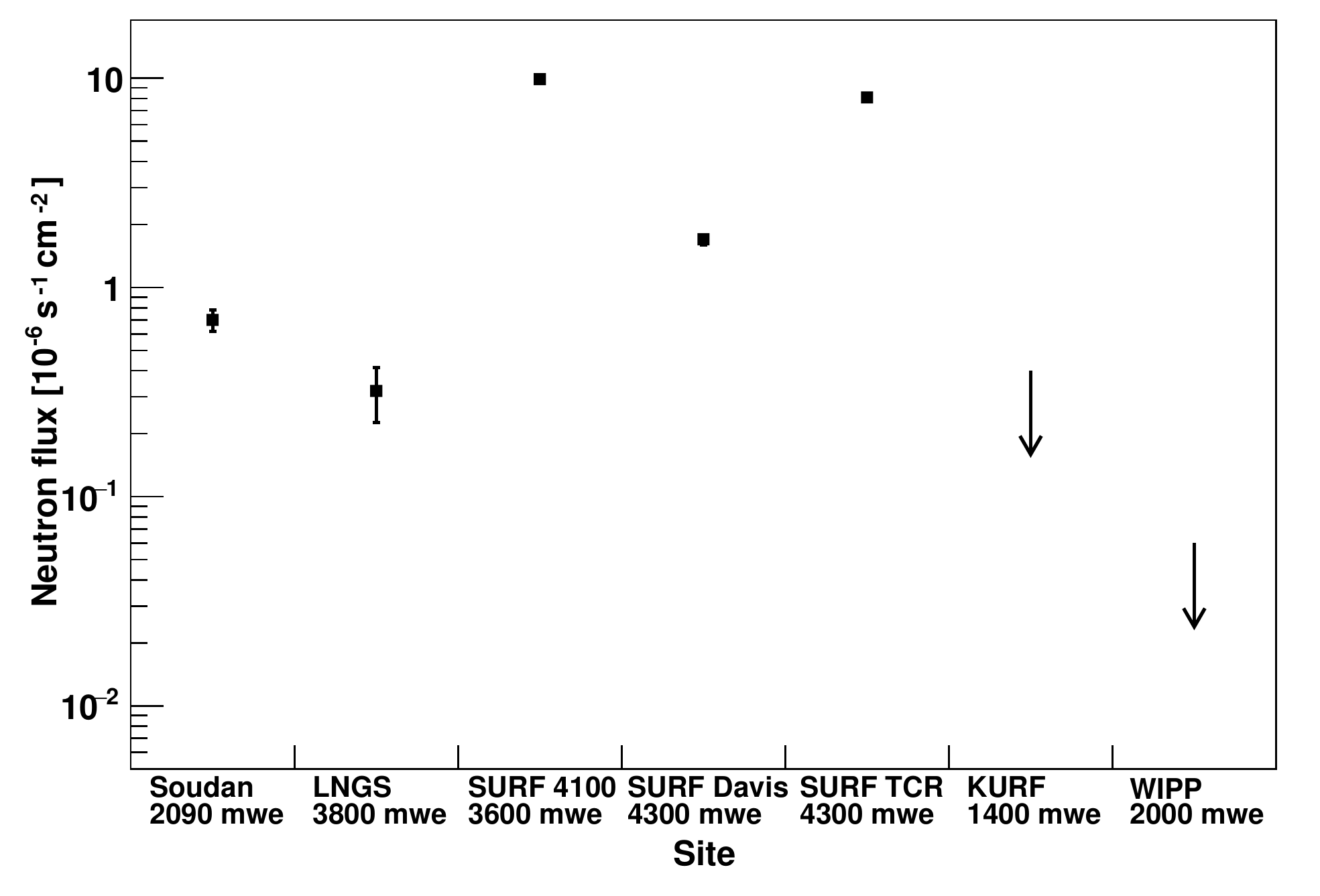}
    \caption{Neutron flux values or upper limits at the visited sites. The average flux on the earth's surface is of the order of $10^{-3}$ s$^{-1}$ cm$^{-2}$.}
    \label{fig:fluxes}
\end{figure}

We presented thermal neutron flux measurements at the main underground laboratories in the United States and at LNGS, presently the worlds largest undergound laboratory.
Since identical equipment was used to measure at each lab this data provides for the first time a systematic comparison of the relative neutron
background levels at the different underground sites. Fluxes above a few $10^{-7}$ s$^{-1}$ cm$^{-2}$ have been determined with reasonable statistics with measurement times of a few weeks; the main 
limiting factor is the intrinsic radioactivity of the counters themselves, which can be clearly seen in the very long WIPP measurement. 

A comparison of the fluxes at each site is shown in Fig. \ref{fig:fluxes}. The measured values lie in the expected ranges and show a reduction in the flux of  3-4 orders of magnitude compared to the surface.
We agree with the lowest reported value from LNGS \cite{Debicki:2009} and reach a comparable uncertainty. The variation in flux between the three locations at SURF should not be seen as very surprising:
there is no reason to expect that the neutron flux at a depth of approximately 1400 m is higher -- or lower, for that matter -- than on a mine level 250 m below. As stated in the introduction, due to the 
large reduction in cosmic rays at deep underground locations the neutron flux is entirely dominated by the radioactive components in the laboratory environments, e.g., the rocks and shotcrete near the experimental
setup. Another factor which can influence the local flux is the radon concentration in the vicinity, which is strongly dependent on the air circulation. In fact the highest neutron flux was measured at SURF's 4100 foot level, where the detectors were situated in a tightly enclosed volume with very little air flow.

As is to be expected from a salt environment WIPP has by far the lowest neutron flux of the visited sites. Our upper limit is lower than the previously reported flux; aside from systematic differences
this can be due to different concentrations of radioactive isotopes near the older measurement location. The local neutron flux is very sensitive to variations in the composition of the surrounding environment.

\section*{Acknowledgments}
This work was funded by the National Science Foundation through grants number PHY-0918728 and PHY-1242506. We acknowledge the hospitality of all underground laboratories involved in this study and express our gratitude for their support: the Kimballton Underground
Research Facility including the management and staff at Lhoist North America - Kimballton; the Soudan Underground Laboratory; the Waste Isolation Pilot Plant; the Sanford
Underground Research Facility; and the LUNA experiment at the Laboratori Nazionali del Gran Sasso. The help of our local contacts at each underground site, Bruce Vogelaar, Jerry Meier, Anthony Villano, Jaret Heise, Norbert Rempe, Steve Elliott and Kevin McIllwee is greatly appreciated.



\bibliographystyle{elsarticle-num} 
\bibliography{underground}





\end{document}